\documentclass[conference]{IEEEtran}
\pdfoutput=1
\usepackage{algpseudocode}
\usepackage{algorithm}

\usepackage{graphicx}
\usepackage{graphics}

\usepackage{url}

\usepackage{subcaption}

\usepackage{cite}
\usepackage{textcomp}

\usepackage{color, colortbl}
\definecolor{green}{rgb}{0.7, 0.93, 0.36}
\definecolor{yellow}{rgb}{0.99, 0.97, 0.37}
\usepackage{multirow}
\usepackage{hhline}

\begin{document}
%
\title{PaREM: A Novel Approach for Parallel Regular Expression Matching}


\author{\IEEEauthorblockN{Suejb Memeti and
		Sabri Pllana}
	\IEEEauthorblockA{Department of Computer Science,
		Linnaeus University\\
		351 95 V\"{a}xj\"{o}, Sweden\\
		\{suejb.memeti, sabri.pllana\}@lnu.se}}

\IEEEspecialpapernotice{\footnotesize(CSE-2014, \copyright IEEE)}

\maketitle

\begin{abstract}
Regular expression matching is essential for many applications, such as finding patterns in text, exploring substrings in large DNA sequences, or lexical analysis. However, sequential regular expression matching may be time-prohibitive for large problem sizes. In this paper, we describe a novel algorithm for parallel regular expression matching via deterministic finite automata. Furthermore, we present our tool PaREM that accepts regular expressions and finite automata as input and automatically generates the corresponding code for our algorithm that is amenable for parallel execution on shared-memory systems. We evaluate our parallel algorithm empirically by comparing it with a commonly used algorithm for sequential regular expression matching. Experiments on a dual-socket shared-memory system with 24 physical cores show speed-ups of up to $21\times$ for 48 threads.  
\end{abstract}

\begin{IEEEkeywords}
	parallel processing, multi-core, regular expression, finite automata
\end{IEEEkeywords}

\IEEEpeerreviewmaketitle

\section{Introduction} \label{introduction}
There are many relevant applications of regular expression matching (REM) and finite automata (FA) including DNA sequence matching \cite{Nowzari-DNA}, network intrusion detection \cite{Babu-NetworkIntrusion}, and information extraction from web based documents\cite{Kosala-informationextraction}. The computational complexity of pattern finding grows with increasing the number of states of the automaton and the size of the input. While the stagnation in processor clock rates promises no performance increases for sequential implementations of REM, availability of affordable multicore processors provides opportunities for significant improvement. For instance, the recently introduced Intel\textregistered~Xeon\textregistered~Processor E7-8890 v2 manufactured at 22nm comprises 15 physical cores and supports 30 threads or so called logical cores. Shared-memory systems with up to eight processors of this type are feasible that would lead to a system with 240 logical cores. To exploit these powerful systems, scalable parallel REM implementations are required.

Programming and resolving problems within automata theory is a relatively complex and time-consuming process, and still the results may not be reliable because of the chances to have an incorrect FA representation. Furthermore, efficient parallel programming of multicore systems is complex and this issue is known in the literature as the \emph{"programmability wall"} \cite{PllanaBMNX08}. Democratization of parallel REM would benefit from tools that hide parallel programming from the end-user and automatically generate the correct parallel implementation that is ready for compilation and efficient execution. 

Various approaches for increasing the performance of REM evaluation have been proposed. For instance, Maine \cite{Maine} is a library for data-parallel FA, which formalizes the evaluation of a FA as a matrix multiplication. Holub and Stekr \cite{HolubS09} propose an algorithm for parallel execution of synchronized deterministic finite automata (DFA). Yang and Prassana \cite{yang2012} introduce an approach that uses segmentation for regular expression evaluation via nondeterministic finite automata (NFA). In \cite{mytkowicz2014} authors propose the range-coalesced representation of transition table to optimize the cost of the transition table lookup for each active state. While there are model to text generators (such as, Acceleo \cite{Acceleo}), or RE to NFA-DFA converters (such as, JFLAP \cite{JFLAP}), to our best knowledge there are no automatic parallel code generators for RE or FA. 

In this paper, we describe a novel algorithm for \underline{Pa}rallel \underline{R}egular \underline{E}xpression \underline{M}atching (PaREM) that scales gracefully for various problem sizes and number of threads. The algorithm was devised to be efficient for general automata independently from the number of states, and for large spectrum of input text-sizes. Our algorithm is optimized to do very accurate speculations on the possible initial states for each of the sub inputs (split among the available processing units), instead of calculating the possible routes considering each state of the automaton as initial state. This method is more effective when the adjacency matrix (used for graph representation of the automaton) is sparse, although it shows major improvements in dense matrices as well. To ease the access to the proposed parallel algorithm for a broad spectrum of users (including the users without background in parallel programming), we have developed our tool PaREM that can transform automatically a Regular Expression (RE) or FA into the corresponding code (C++ and OpenMP) for our algorithm that is amenable for parallel execution on shared-memory systems.
Experimental results on a dual-socket shared-memory system with 24 physical cores show a close to linear speedup compared to the sequential implementation for problem sizes comparable to the cache size and significant speedup for larger problem sizes that use further levels of memory hierarchy.

The main contributions of this paper include:
\begin{itemize}
	\item A scalable algorithm for parallel regular expression matching;
	\item PaREM tool that automatically generates parallel code from a given regular expression or finite automata;
	\item Empirical evaluation of the proposed parallel algorithm and the PaREM tool using a modern dual-socket shared-memory system with 24 physical cores.
\end{itemize}

The rest of the paper is organized as follows. Section \ref{methodology} provides background information on regular expressions and finite automata and presents our parallel algorithm. Section \ref{implementation} describes the implementation of the PaREM tool, and Section \ref{exp_evaluation} the corresponding experimental evaluation. The work described in this paper is compared and contrasted to the related work in Section \ref{related_work}. Section \ref{summary_future_work} provides a summary of our work and a description of future work.

\section{Methodology} \label{methodology}

\subsection{Background}
A regular expression is a string for describing search patterns. A finite automaton is a graph-based way for specifying patterns \cite{aho1994foundations}. Finite automata and regular expressions may be used in pattern finding algorithms.

Deterministic Finite Automata (DFA) is a quintuple of $(Q, \Sigma,\delta, q_{0}, F)$, where $Q$ is a finite set of states, $\Sigma$ is set of symbols (alphabet), $\delta: Q \times \Sigma\rightarrow Q$ is the transition function, $q_{0}$ is the initial state and $F$ is the set of final states \cite{aho1994foundations}\cite{hopcroftullman}. A DFA operates in the following manner: when a program starts, the current state is assumed to be the initial state $q_{0}$, on each character the current symbol is supposed to move to another state (including itself). When the input reaches the last character, the string is accepted if and only if the current state is in the set of final states. It is called deterministic because in each state and for each input symbol a unique transition is defined. 

Nondeterministic Finite Automata (NFA) is defined by the quintuple $(Q, \Sigma, \delta, q_{0}, F)$ as in DFA except the alphabet may contain an empty symbol; the transition function returns a set of states rather than a single state. It is called non-deterministic because of the choice of moves that may lead from one state to another.

\subsection{Parallel REM Algorithm}
Existing approaches for parallel REM (such as \cite{HolubS09}) split the input into smaller substrings among all or a selected number of processing units, run the automaton on each of them, and join the sub- results. While other approaches calculate the possible initial states from each state of the automaton, our algorithm takes a step ahead by excluding all the states that the automaton has no outgoing or incoming transitions for the specified characters. Calculating the possible routes from each state of the automaton becomes time-consuming and memory-expensive for large finite automatons.  

The basic idea of the sequential REM or DFA is that one starts from $q_{0}$ and after n (input length) steps another state from set $Q$ is reached. Its time complexity depends only on the input length.

Our algorithm is based on domain decomposition, which means it slices the input in $p$ parts (see Algorithm \ref{alg_parallel}), where $p$ is the number of processing units (line 3). For each $p_{i}$ the possible initial states $R$ are determined by finding the intersection of possible initial states $R = S \cap L$ (line 5 --- 15). $S$ is the set of initial states for the first character of $T_{p_{i}}$ (that is, the sliced input for this specific processor) where $q_{i} \in S$ if $\exists:\delta(q_{i}, T_{p_{i}})$. $L$ is the set of initial states for last character of $T_{p-1_{i}}$, where $L_{i} = \delta(q_{i}$, $T_{p-1_{i}})$.
Each chunk of the input is mapped to a processing unit, and each processing unit is responsible for finding the possible initial states for its own chunk of the input. The processing unit with $ID = 0$ already knows the possible initial state, that is $q_{0}$, so a calculation for determining the possible initial states is not necessary. For each state in $R$, a REM is done and the result is stored in $I$ (lines 16 --- 25). 

When all processors have finished their jobs, a binary reduction of the final results is completed. The reduction is done by connecting the last active state of $P_{i}$ to the first active state of $P_{i-1}$. The connection is accepted only if a transition from last active state of $P_{i}$ to the first active state of $P_{i+1}$ exists with the first character of the sub-result of next processor $T_{p+1}$,$\delta (q_{i}, T_{p+1_{i}})$. An input is accepted only if for each processor there exist a sub route, which can be connected with the result of the previous and next processor's result, and the last state of the automaton is member of the final state set. The worst-case scenario would be if all the states have the same input and output transitions. 

\begin{algorithm}
	\caption{Parallel Regular Expression Matching (PaREM)}
	
	\begin{algorithmic}[1]
		\Statex { \%Input: Transition table Tt, set of final states F, input T\%}
		\Statex { \%Output: Result of REM\%}
		\State {$I = vector(p)$ /* initialize final result vector */ }
		\Statex { \%$P_{0} ... P_{p}...$ processing unit, $p$ is the total number of processing units \% }
		\For  {$P_{0}, P_{1}, ..., P_{p} $} in parallel
		\State $start\_position = i * (T.length / p) $
		\State $pi\_input = substring(start\_position, T.length / p) $
		\Statex { \%start find possible initial states \%}
		\For  {$q_{0}, q_{1}, ..., q_{n} $} 
		\Statex { \% $pi\_input.at(0)$ returns the first char of $pi\_input$ \% }
		\If{  $(Tt[q_{i}][pi\_input.at(0)]\in Q)$ }
		\State $S[i] = q_{i}$
		\EndIf 
		\EndFor
		
		\For  {$q_{0}, q_{1}, ..., q_{n} $} 
		\Statex { \% $pi\_input.back()$ returns the last char of $pi\_input$ \% }
		\If{  $(Tt[q_{i}][pi\_input.back()]\in Q)$ }
		\State {$L[i] = Tt[q_{i}][pi\_input.back()]$}
		\EndIf 
		\EndFor
		\Statex { \%end find possible initial states \%}
		
		\State $R = S \cap L$ \%intersection of possible initial and last states \%
		
		\For {$r \in R$} 
		\State $Rr = vector(pi\_input.length())$ 
		\For {$char \in pi\_input$} 
		\If{  $(Tt[r][char] \in F)$ }
		\State $found++$
		\EndIf 
		\State $Rr[i] = r = Tt[r][char]$ 
		\EndFor
		\State $I[i].push\_back(Rr)$
		\EndFor

		\EndFor		
		
		\Statex{ \% Wait for the slowest processor\%} 
		\Statex{ \% Perform a reduction of I\%} 
		
	\end{algorithmic}
	\label{alg_parallel}
\end{algorithm}

\subsection{Description of PaREM Algorithm with an Example}

\begin{figure}[!t]
\centering
\includegraphics[width=3in]{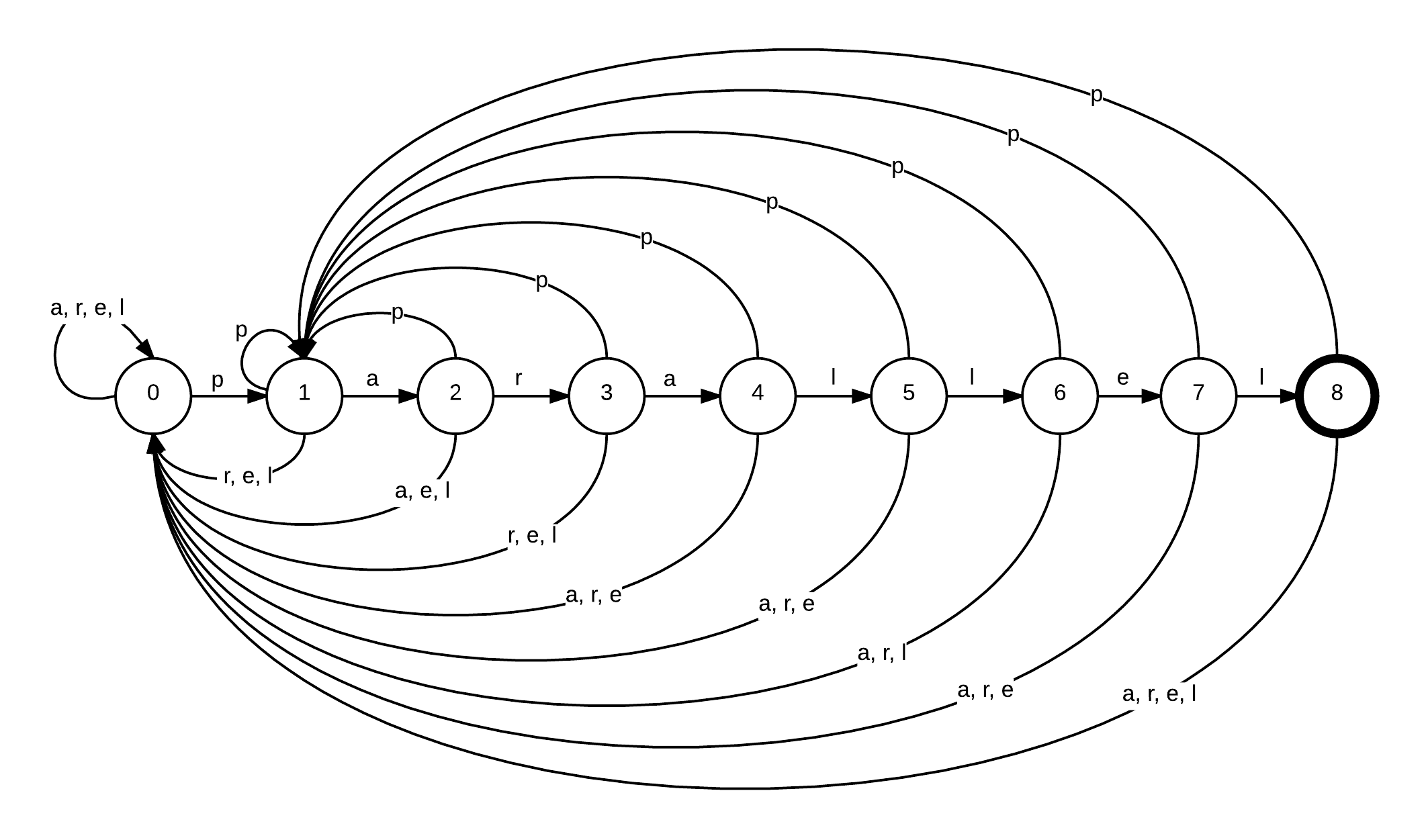}
\caption{Automaton $A$ for matching the pattern \textit{parallel}}
\label{fig_automaton}
\end{figure}

To show how the possible initial states are determined, the following example from Fig. \ref{fig_automaton} is used. Let $T$ be an input string, $T = "plaraparallelapareparapl"$ and assume that we will use four processing units (that is threads).

The transition table corresponding to the automaton from Fig. \ref{fig_automaton} is shown on Table I. The transition table for this automaton is dense, which will produce a dense adjacency matrix.

\begin{table}[!t]
\renewcommand{\arraystretch}{1.3}
\caption{Transition table for automaton on Fig. \ref{fig_automaton}}
\label{transition_table}
\centering
\begin{tabular}{|p{0.5cm}|p{0.5cm}|p{0.5cm}|p{0.5cm}|p{0.5cm}|p{0.5cm}|}
\hline 
$\delta_{A}$ & p & a & r & e & l \\
\hline
0 \cellcolor{green} & 1 & 0 & 0 & 0 & 0  \\
\hline
1 \cellcolor{green} & 1 & 2 & 0 & 0 & 0 \\
\hline
2 \cellcolor{green} & 1 & 0 & 3 & 0 & 0 \\
\hline
3 \cellcolor{green} & 1 & 4 & 0 & 0 & 0 \\
\hline
4 \cellcolor{green} & 1 & 0 & 0 & 0 & 5 \\
\hline
5 \cellcolor{green} & 1 & 0 & 0 & 0 \cellcolor{yellow} & 6 \\
\hline
6 \cellcolor{green} & 1 & 0 & 0 & 7 \cellcolor{yellow} & 0 \\
\hline
7 \cellcolor{green} & 1 & 0 & 0 & 0 & 8 \\
\hline
8 \cellcolor{green} & 1 & 0 & 0 & 0 & 0 \\
\hline
\end{tabular}
\end{table}

The input length is 24 characters, so when split among processing units we get four substrings of six characters ($P_{0} = "plarap"$, $P_{1} = "aralle"$, $P_{2} = "lapare"$ and $P_{3} = "parapl"$). Table \ref{possible_initial_states} shows the accurate possible initial states found for each of the processor's input, and the visited states starting from each of the possible initial states. In this example, each state has exactly the same amount of outgoing transitions, which means there is a transition from each state for each symbol of the alphabet. 

The set of DFA initial states R is equal to the set of states $L$ achieved from the last character of the input string of the previous processor, because $S$ is equal to set of all states. Therefore, $R = S \cap L = L$. This applies only to dense transition tables, because from each state on any symbol is possible to go to another state (including itself). In practice, most of DFA produce a sparse transition table. In sparse transition tables the set of states $S$ achieved from the first character of the input string that is mapped to the processing unit, is determined by the outgoing transitions of states for a specific character. We treat each matrix as sparse, that is why $R = S \cap L$. It is possible to identify a sparse matrix, but inspecting each element of large matrices whether is empty or not may be time-consuming. 

\begin{table}[!t]
	\renewcommand{\arraystretch}{1.3}
	\caption{Possible initial states for $P_{0}, P_{1}, P_{2}~and~P_{3}$}
	\label{possible_initial_states}
	\centering
	\begin{tabular}{|p{0.5cm}|p{1cm}|p{2cm}|}
		\hline 
		{}  & $S \cap L$ & Visited states\\
		\hline
		$P_{0}$ & 0 & 1 0 0 0 0 1 \\
		\hline
		$P_{1}$ & 1 & 2 3 4 5 6 7 \\
		\hline
		\multirow{2}{*}{$P_{2}$} & 0 & 0 0 1 2 3 0\\
		{}&7& 8 0 1 2 3 0\\
		\hline
		
		$P_{4}$ & 0 & 1 2 3 4 1 0 \\
		
		\hline
	\end{tabular}
\end{table}

The highlighted numbers on Table \ref{transition_table} represent the set of states $S$ and $L$ for $P_{2}$, where $S$ (colored in green) is set of source states for which a transition exist on $"l"$ (first character of the input mapped to $P_{2}$), and $L$ (colored in yellow) is set of unique destination states for which a transition exists on $"e"$ (last character of the input string mapped to $P_{1}$).

The general enumeration approach of REM algorithms calculates possible routes (moving from one state to another) considering each state of the automaton as initial state. In this example, the enumeration approach of REM would have performed $3\times9+1$ (three processing units ($P_{1}, P_{2}$ and $P_{3}$) would start from all the nine possible states, and $P_{0}$ would start from state $q_{0}$) calculations. Our algorithm performs only five calculations for this example, and we believe that this number becomes lower for sparse transition tables. If the input of processing unit $P_{i-1}$ would end with $"l"$, there would be four (0, 5, 6, 8) possible initial states. The worst-case scenario would be if each of the sub-inputs ends with $"l"$; in such case $3\times4+1$ calculations are performed for dense matrices that is an improvement by $2.15 = (3*9+1) / (3*4+1)$, compared to the general approach.

\section{Implementation} \label{implementation}
\begin{figure*}[!t]
	\centering
	\includegraphics[width=\textwidth]{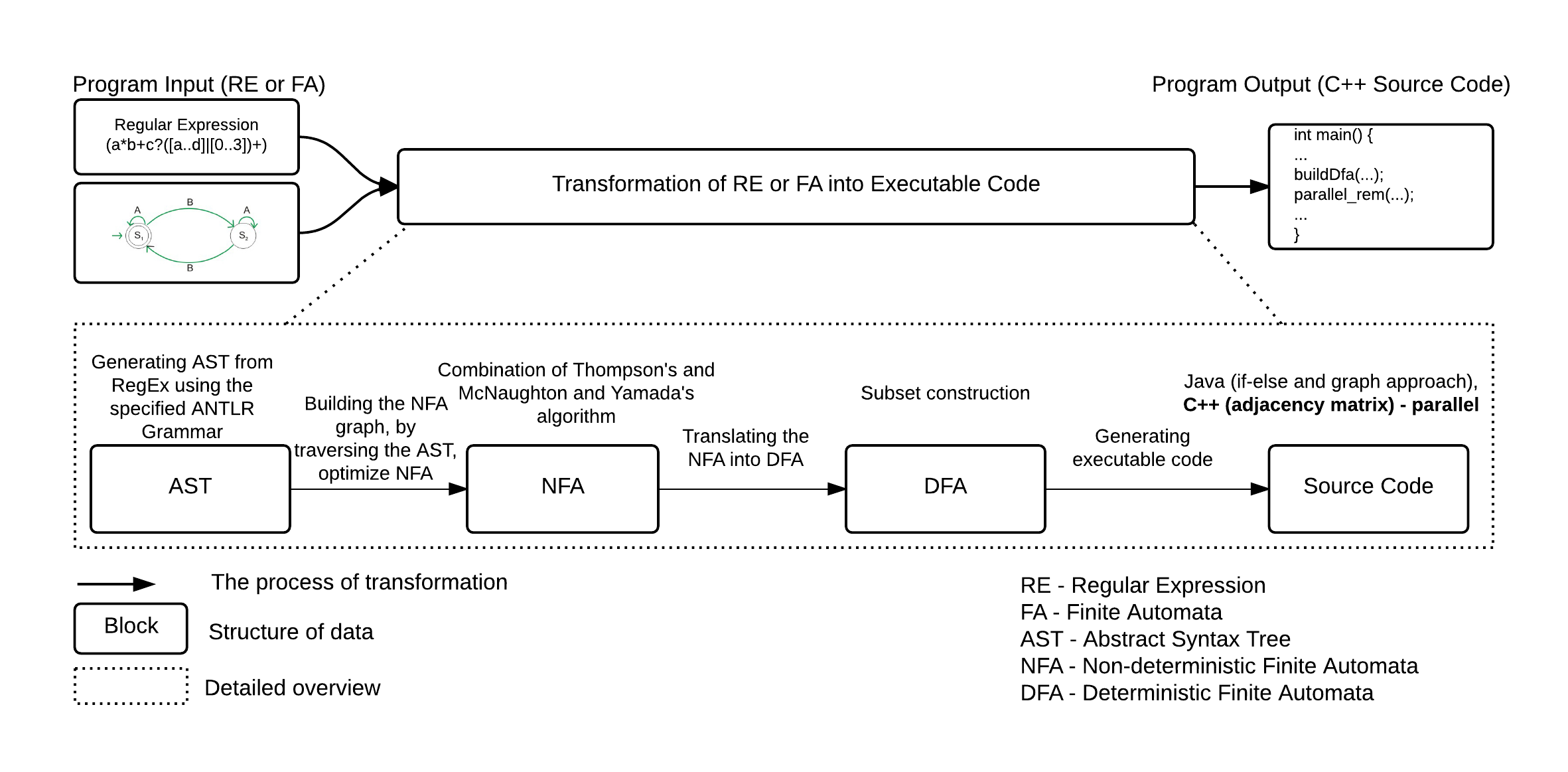}
	\caption{The use of PaREM tool for translating regular expressions into equivalent finite automata (NFA, then DFA) and generating source code (C++ and OpenMP) that represent the same given RE or FA}
	\label{fig_toolchain}
\end{figure*}

Fig. \ref{fig_toolchain} depicts our PaREM tool, which takes as input a RE or a FA and generates the corresponding C++ code representation of the given RE or FA. The generated C++ code includes OpenMP \cite{OpenMP} directives and routines and is in accordance with our Algorithm \ref{alg_parallel}.
In the process of PaREM implementation, we have specified a context-free grammar to define the language that accepts regular expressions as input. Table \ref{operators_table} lists the accepted operators by PaREM context-free language. 

\begin{figure}[!t] 
	\label{ fig_nfatrans}
	\begin{subfigure}[b]{0.45\linewidth}
		\centering
		\includegraphics[width=\linewidth]{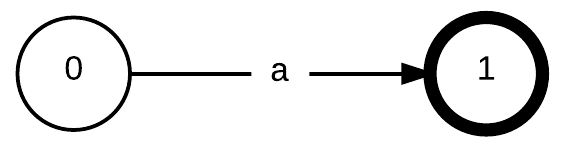}
		\caption{RE input: $a$}
		\label{nfa:a}
		\vspace{4ex} 
	\end{subfigure} 
	\begin{subfigure}[b]{0.45\linewidth}
		\centering
		\includegraphics[width=\linewidth]{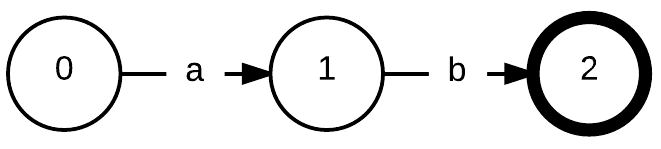}
		\caption{RE input: $ab$}
		\label{nfa:b}
		\vspace{4ex}
	\end{subfigure} 
	\begin{subfigure}[b]{0.45\linewidth}
		\centering
		\includegraphics[width=\linewidth]{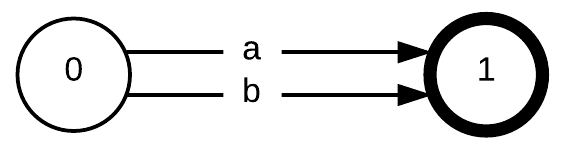}
		\caption{RE input: $a|b$} 
		\label{nfa:c}
		\vspace{4ex}
	\end{subfigure}
	\hfill
	\begin{subfigure}[b]{0.45\linewidth}
		\centering
		\includegraphics[width=\linewidth]{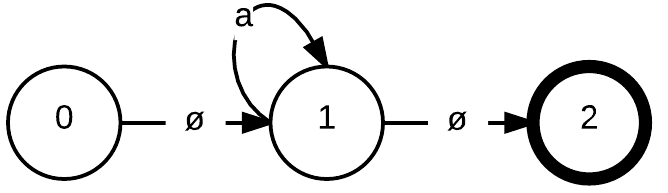}
		\caption{RE input: $a*$} 
		\label{nfa:d}
		\vspace{4ex}
	\end{subfigure} 
	\begin{subfigure}[b]{0.45\linewidth}
		\centering
		\includegraphics[width=\linewidth]{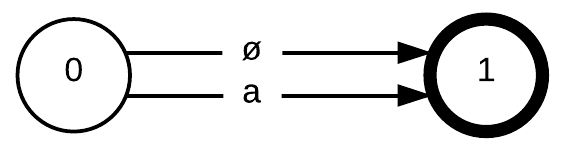}
		\caption{RE input: $a?$} 
		\label{nfa:e}
		\vspace{4ex}
	\end{subfigure}
	\hfill
	\begin{subfigure}[b]{0.45\linewidth}
		\centering
		\includegraphics[width=\linewidth]{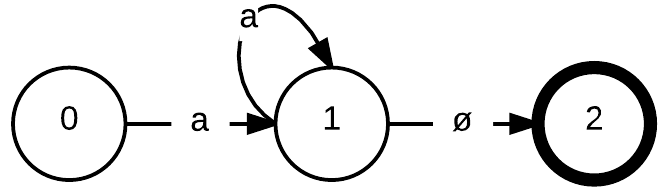}
		\caption{RE input: $a+$} 
		\label{nfa:f}
		\vspace{4ex}
	\end{subfigure} 
	\begin{subfigure}[b]{\linewidth}
		\centering
		\includegraphics[width=\linewidth]{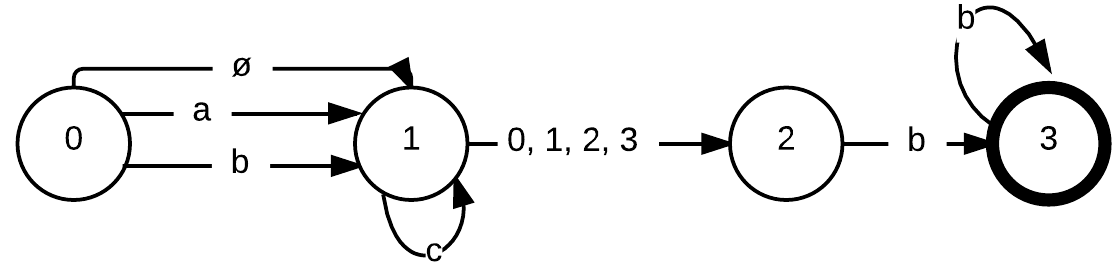}
		\caption{RE input: $(a|b)?c*[0..3]b+$} 
		\label{nfa:g}
		\vspace{4ex}
	\end{subfigure}
	\caption {Transformation of RE operators into NFA}
\end{figure}

The \textit{Klenee Star} denotes zero or more occurrences of a symbol or sub-expression (for instance, $\phi, a, aa, aaa$, where $\phi$ is an empty transition). The NFA representation of the \textit{Klenee Star} is shown in Fig. \ref{nfa:d}. The \textit{Positive Closure} also known as \textit{Repetition} is an extended operator of the \textit{Klenee Star}, which denotes one or more occurrences of a symbol or sub-expression (for instance, $a+$, Fig. \ref{nfa:f}, is equal to $aa*$ that results to these possibilities: $a, aa, aaa, …$).

The \textit{Union} operator (represented as NFA in Fig. \ref{nfa:c}), expressed by a vertical bar, provides the possibility to choose between two or more sub-expressions (such as, $a, b$). The \textit{Range} (defined based on ASCII code order) operator, or \textit{Character Class}, is an extended operator of \textit{Union}, instead of writing $0|1|2|3$ the \textit{Range} operator $[0..3]$ can be used. It applies to integers and characters.

The \textit{Optionality} operator (shown as NFA in Fig. \ref{nfa:e}) denotes zero or one occurrence of a symbol or sub-expression (for instance, $a? = \phi|a$). The \textit{Group} operator is introduced to change the operator precedences. For instance, $a|b*$ and $(a|b)*$ produce different results, in the first example the \textit{Klenee Star} operator has priority over the \textit{Union} operator, while in the second example the \textit{Union} operator has a higher priority. By combining these operations (using \textit{Concatenation} operator, Fig. \ref{nfa:b}) arbitrarily complex regular expressions can be written.

\begin{table}[!t]
	\renewcommand{\arraystretch}{1.3}
	\caption{PaREM's Accepted Regular Expressions Operators}
	\label{operators_table}
	\centering
	\begin{tabular}{|p{1cm}|p{2cm}|p{2cm}|}
		\hline 
		Operator & Name & Description\\
		\hline
		$ab$ & Concatenation & $b$ right after $a$\\
		\hline
		$a*$ & Klenee Star & zero or more $a$'s\\
		\hline
		$a|b$ & Union & either $a$ or $b$\\
		\hline
		$a+$ & Positive closure & one or more $a$'s\\
		\hline
		$[0..9]$ & Range & either $0, 1 ...$ or $9$\\
		\hline
		$a?$ & Optionality & zero or one $a$\\
		\hline
		$(ab | c)*$ & Group & zero or more of either $ab$'s or $c$'s\\
		\hline
	\end{tabular}
\end{table}

\begin{figure}[!t]
	\centering
	\includegraphics[width=\columnwidth]{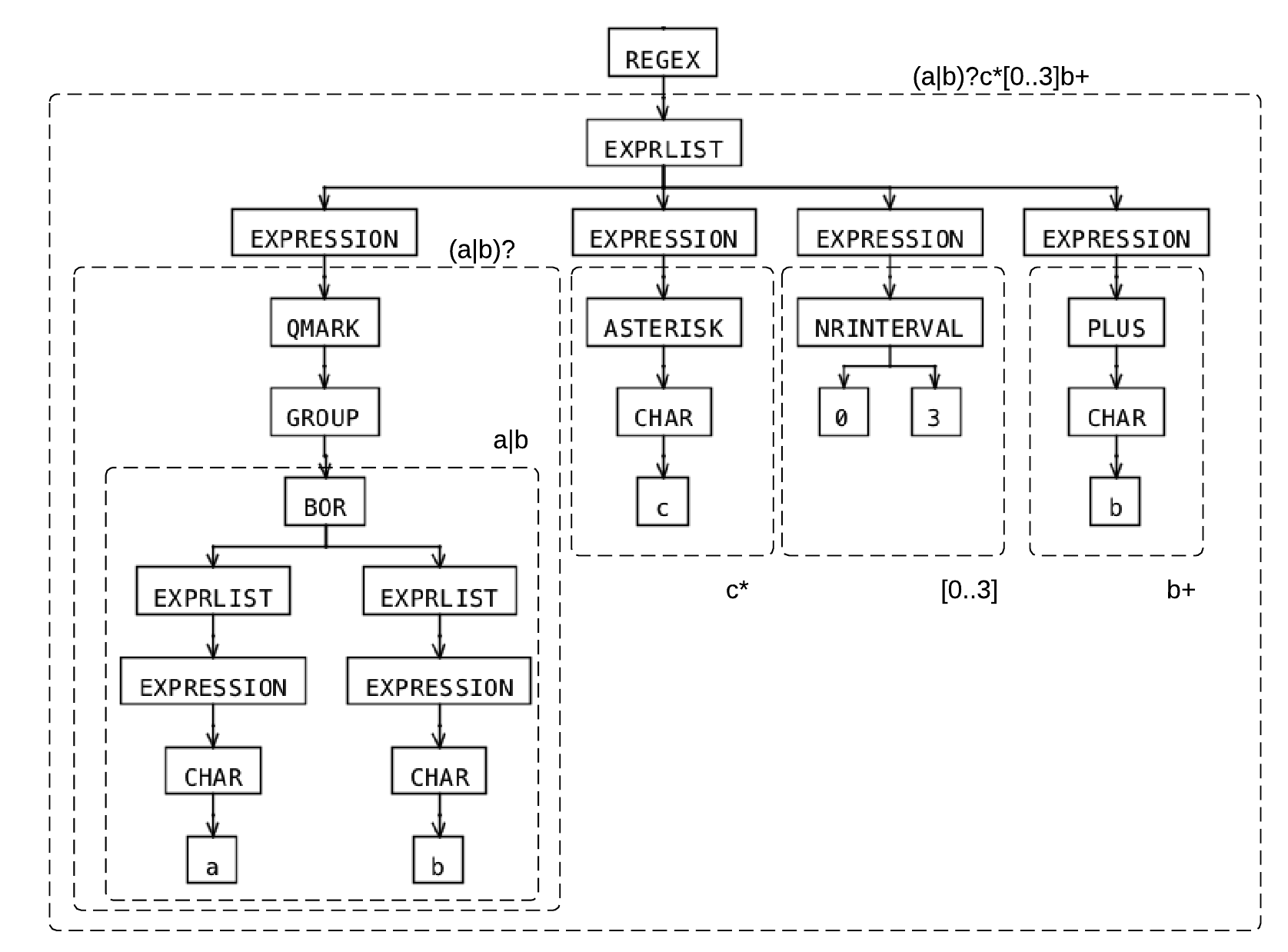}
	\caption{Abstract Syntax Tree representation for $(a|b)?c*[0..3]b+$ RE}
	\label{fig_ast}
\end{figure}

For each RE a specific Abstract Syntax Tree (AST) is generated that represents the abstract syntactic structure of the RE. For easier translation into a target structure, additional details have been added (such as, the node type) to the AST. The generated AST can have an arbitrary number of sub-trees, which in essence are AST’s \cite{ahocompilers}. Fig. \ref{fig_ast} shows an example of how an AST is constructed for a given RE. Dashed-line compartments indicate the sub-trees. 

The priority of the \textit{Union} operator over the \textit{Quantifier} operator in the sub-expression $"(a|b)?"$ is depicted in Fig. \ref{fig_ast}. The deeper the operator is in the AST hierarchy, the higher priority it has. 

We transform the AST into NFA graph using the \textit{McNaughton-Yamada-Thompson Algorithm}. To preserve the operator priority, the depth-first search traversal of the tree is performed while constructing the NFA graph. Each of the sub-expressions creates a sub-graph, which are merged in the main graph using empty transitions. Removing the unnecessary empty transitions further optimizes the final NFA. The optimized NFA for the RE example in Fig. \ref{fig_ast} is shown in Fig. \ref{nfa:g}.

Fig. \ref{nfa:a} --- \ref{nfa:f} depicts the transformation process for each operator from the RE (or AST) into an equivalent NFA. 

Using the \textit{Subset Construction Algorithm} \cite{ChangP97}, the optimized NFA is converted into an equivalent DFA.  During this transformation, the PaREM creates a log file with the transition table. Theoretically, the DFA’s number of states may have an exponential relationship to the NFA’s number of states, which leads to the well-known state explosion issue. However, most of the real-world NFA produce a DFA with approximately the same number of states. 

Finally, from the DFA we generate executable source code that implements the REM for the corresponding DFA \cite{ahocompilers}\cite{arora2005comprehensive}. There are different possible ways of representing a DFA, but we have selected two different forms: (1) \textit{if-else} statements, and (2) \textit{graphs}. 

The \textit{if-else} approach is a straightforward way of implementing a DFA. This approach creates an if-statement for each transition of the automaton. However, this approach is not recommended for large automatons. The \textit{if-else} approach provides a sequential solution for regular expression matching.

The \textit{graph-based} approach provides an easy way to add/remove transitions or states in the automaton, and consequently reduces the risk of having incorrect representation of the automaton.

For \textit{graph-based} representation in the source code, we have used an \textit{adjacency matrix}, which represents the transition table. This approach has faster lookups to check for the presence or absence of a specific transition, compared to the \textit{adjacency list} representation of the automaton. The \textit{graph-based} solution provides the implementation of the parallel regular expression-matching algorithm presented in this paper.

\section{Experimental Evaluation} \label{exp_evaluation}

For experimental purposes, an automaton that finds all occurrences of the word \textit{"parallel"} has been implemented, which results with an automaton with nine states (shown on Fig. \ref{fig_automaton}) and an alphabet of five characters. Table \ref{system_configuration_table} lists the major features of experimentation platform. We use a shared-memory system with two 12-core Intel\textregistered~Xeon\textregistered~processors of the type E5-2695 v2 for evaluation of our approach. Each of the 12 physical cores supports two threads (also known as logical cores). In total, our system has 24 physical cores or 48 logical cores. 

\begin{figure*}[!t] 
	\label{fig_performance}
	\begin{subfigure}[b]{0.45\linewidth}
		\centering
		\includegraphics[width=\linewidth]{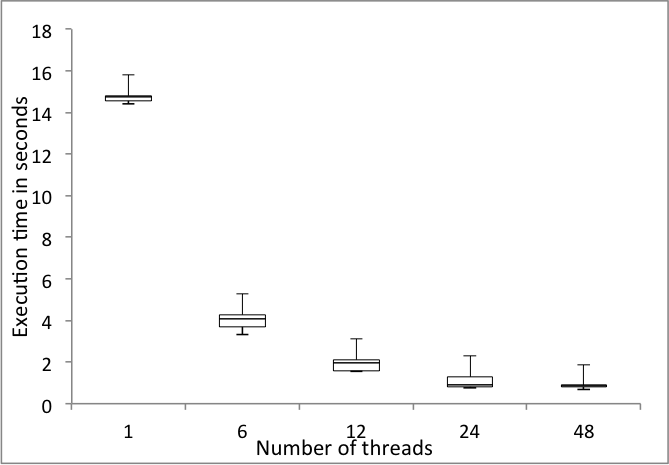}
		\caption{input length: 6.69e+07}
		\label{perf:a}
		\vspace{4ex} 
	\end{subfigure} 
	\begin{subfigure}[b]{0.45\linewidth}
		\centering
		\includegraphics[width=\linewidth]{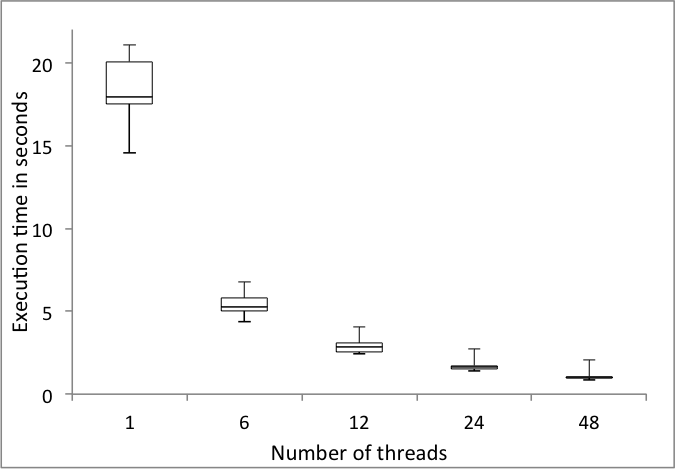}
		\caption{input length: 1.34e+08}
		\label{perf:b}
		\vspace{4ex}
	\end{subfigure} 
	\begin{subfigure}[b]{0.45\linewidth}
		\centering
		\includegraphics[width=\linewidth]{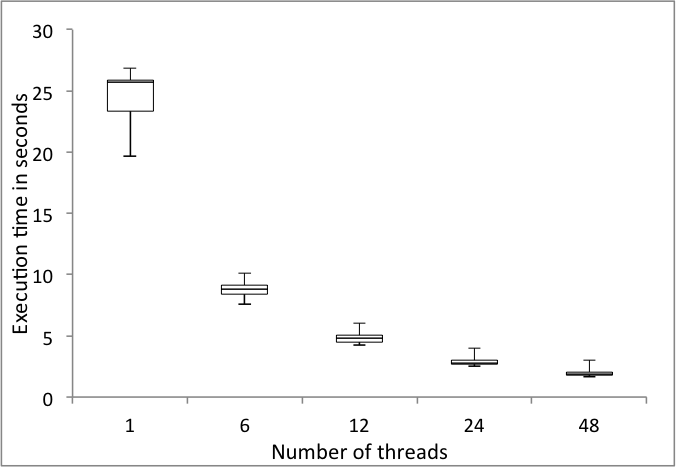}
		\caption{input length: 2.68e+08} 
		\label{perf:c}
		\vspace{4ex}
	\end{subfigure}
	\hfill
	\begin{subfigure}[b]{0.45\linewidth}
		\centering
		\includegraphics[width=\linewidth]{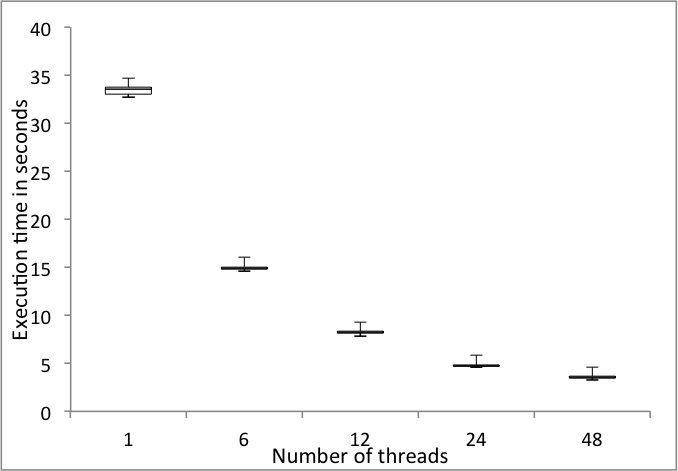}
		\caption{input length: 5.36e+08} 
		\label{perf:d}
		\vspace{4ex}
	\end{subfigure} 
	\begin{subfigure}[b]{0.45\linewidth}
		\centering
		\includegraphics[width=\linewidth]{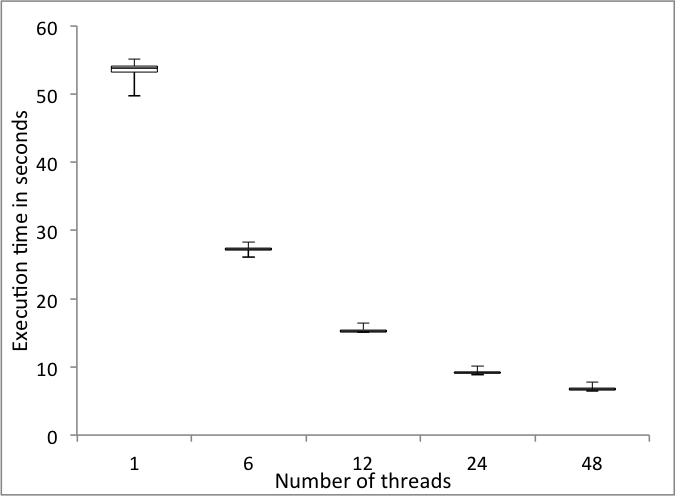}
		\caption{input length: 1.07e+09} 
		\label{perf:e}
		\vspace{4ex}
	\end{subfigure}
	\hfill
	\begin{subfigure}[b]{0.45\linewidth}
		\centering
		\includegraphics[width=\linewidth]{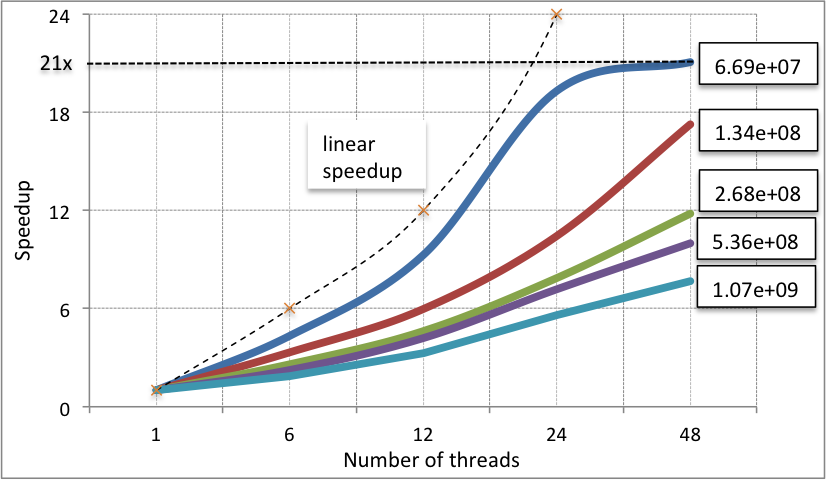}
		\caption{speedup} 
		\label{perf:f}
		\vspace{4ex}
	\end{subfigure} 
	\caption {Performance results. As input are used five strings of the following lengths: 6.69e+07, 1.34e+08, 2.68e+08, 5.36e+08 and 1.07e+09. Execution times are shown in (a – e), whereas the speedup is shown in (f). The speed up for the smallest input length (6.69e+07 characters) in our set of experiments closely follows the linear speedup up to 24 threads. The maximum speedup of $21\times$ is achieved for 48 threads and input string of 6.69e+07 characters.}
\end{figure*}

Fig. \ref{perf:a} --- \ref{perf:e} depicts the performance results for five problem sizes and various numbers of threads. Each experiment has been repeated 20 times to address the random performance fluctuations. The string length determines the problem size and in our experiment, we used five strings of following lengths: 6.69e+07, 1.34e+08, 2.68e+08, 5.36e+08 and 1.07e+09. 

\begin{table}[!t]
	\renewcommand{\arraystretch}{1.3}
	\caption{System Configuration}
	\label{system_configuration_table}
	\centering
	\begin{tabular}{|p{2cm}|p{3cm}|}
		\hline 
		Operating System & CentOS 6.2 (Linux kernel 2.6.32)\\
		\hline
		Processor & $2 \times$ Intel\textregistered~Xeon\textregistered~Processor E5-2695 v2  (2.40GHz, 30MB Cache, 12 Cores)\\
		\hline
		RAM & $8 \times 16$GB\\
		\hline
		OpenMP & 3.1\\
		\hline
	\end{tabular}
\end{table}

Execution times are shown in Fig. \ref{perf:a} --- \ref{perf:e}, whereas the speedup is depicted in Fig. \ref{perf:f}. The speed up for the smallest input length (6.69e+07 characters) in our set of experiments closely follows the linear speedup up to 24 threads (Fig. \ref{perf:f}). For larger input lengths, we may observe noteworthy speedup improvements for 24 and 48 threads. Considering all experiments the highest speedup of $21\times$ was achieved for input length 6.69e+07 characters and 48 threads. 

Table \ref{cache_misses_table} shows the influence of input length in the cache misses and the speedup. We varied the input length using 24 and 48 threads. With the increase of input length, the number of cache misses increases and the speedup decreases. For the smallest input length in our set of experiments (6.69e+07 characters) that largely fits in the available cache, using 24 threads, the number of cache misses is 36.34e+06 and the speedup is $19.32\times$. For the largest input length (1.07e+09) we obtained 681.71e+06 cache misses and a speedup of $5.62\times$.

The obtained cache misses for 48 threads are comparable to those for 24 threads (see Table \ref{cache_misses_table}). For the smallest input length the number of cache misses is 36.76e+06 and the speedup is $21.08\times$. For the largest input length (1.07e+09) we obtained 716.02e+06 cache misses and a speedup of $6.69\times$. We may observe that for all tested input lengths the speedup-gain when 48 logical cores (hyper-threading) are used compared to 24 physical cores.

\begin{table}[!t]
	\renewcommand{\arraystretch}{1.3}
	\caption{Influence of Input length in cache misses and speedup for 24 and 48 threads}
	\label{cache_misses_table}
	\centering
	\begin{tabular}{|p{1.4cm}|p{1.2cm}|p{1.2cm}|p{1.2cm}|p{1.2cm}|}
		\hline 
			\multirow{2}{*}{Input Length} & \multicolumn{2}{c|}{24 threads} & \multicolumn{2}{c|}{48 threads}\\
			\cline{2-5}
			{} & Cache Misses [106] & Speedup & Cache Misses [106] & Speedup\\
			\cline{2-5}
		\hline
		{6.69e+07} & 36.34 & 19.32 & 36.76 & 21.08\\
		\hline
		{1.34e+07} & 70.15 & 10.44 & 71.07 & 17.27\\
		\hline
		{2.68e+08} & 167.57 & 7.87 & 140.57 & 11.81\\
		\hline
		{5.36e+08} & 339.26 & 7.18 & 367.07 & 9.99\\
		\hline
		{1.07e+09} & 681.71 & 5.62 & 716.02 & 6.69\\	
		\hline
		
	\end{tabular}
\end{table}

\subsection{Performance comparison of PaREM algorithm with the General Enumeration Approach}
The main difference of the PaREM algorithm and the General Enumeration Approach (Enum) proposed by \cite{HolubS09} is the way of speculation of the next set of possible initial states for each chunk of the input string. While the Enum algorithm for general DFAs considers all the states of the automaton as initial states, the PaREM algorithm finds the most accurate initial states. Comparing to PaREM that requires only five calculations to find the correct path, the Enum algorithm requires 28 calculations to be performed in order to find the correct initial states for the example described in section \ref{methodology}.C. 

\begin{figure}[!t] 
	\label{fig_comparison_parem_vs_enum}
	\begin{subfigure}[b]{\linewidth}
		\centering
		\includegraphics[width=0.77\linewidth]{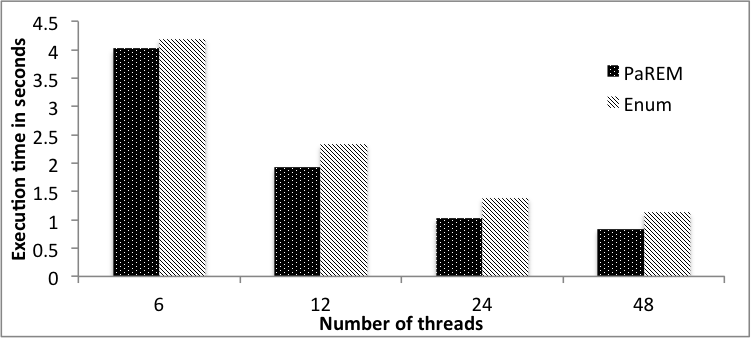}
		\caption{input length: 6.69e+07}
		\label{comparison:a}
		\vspace{4ex} 
	\end{subfigure} 
	\begin{subfigure}[b]{\linewidth}
		\centering
		\includegraphics[width=0.77\linewidth]{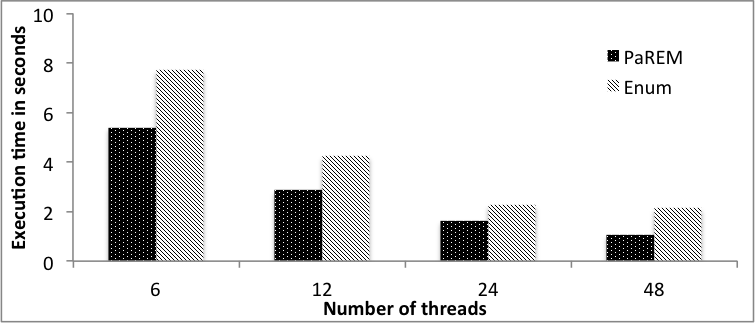}
		\caption{input length: 1.34e+08}
		\label{comparison:b}
		\vspace{4ex}
	\end{subfigure} 
	\begin{subfigure}[b]{\linewidth}
		\centering
		\includegraphics[width=0.77\linewidth]{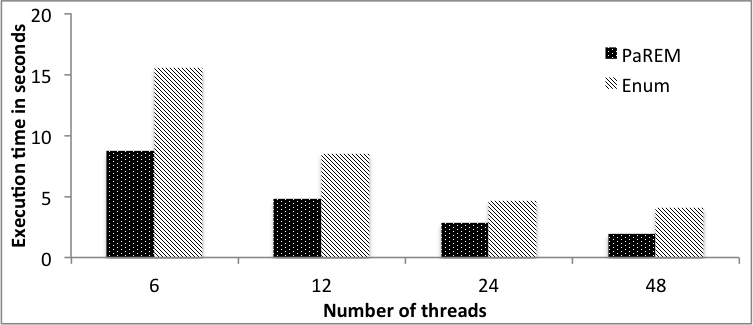}
		\caption{input length: 2.68e+08} 
		\label{comparison:c}
		\vspace{4ex}
	\end{subfigure}
	\begin{subfigure}[b]{\linewidth}
		\centering
		\includegraphics[width=0.77\linewidth]{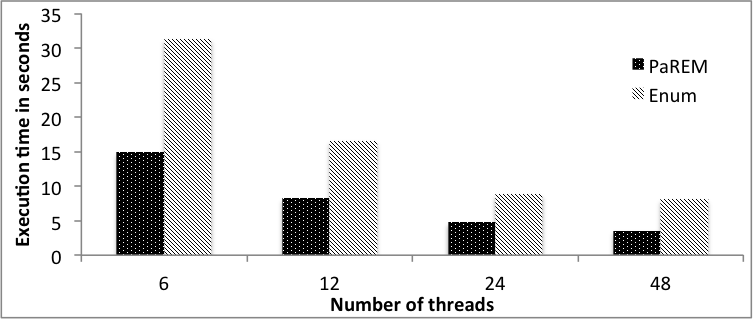}
		\caption{input length: 5.36e+08} 
		\label{comparison:d}
		\vspace{4ex}
	\end{subfigure} 
	\begin{subfigure}[b]{\linewidth}
		\centering
		\includegraphics[width=0.77\linewidth]{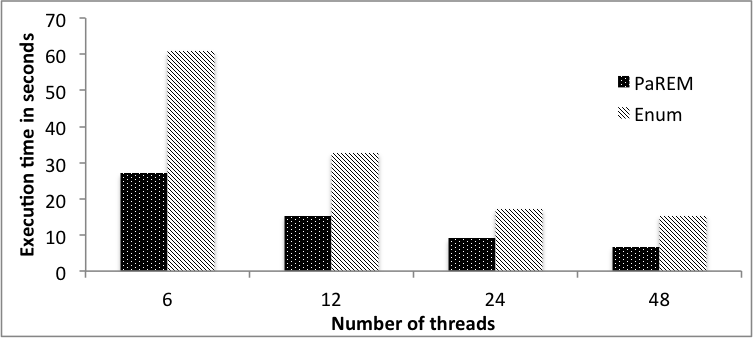}
		\caption{input length: 1.07e+09} 
		\label{comparison:e}
		\vspace{4ex}
	\end{subfigure}
	
	\caption {Comparison between PaREM algorithm and General Enumeration Approach.}
\end{figure}

We have run the experiment example from section \ref{methodology}.C with the same input sizes and number of threads for the General Enumeration approach as well. Figure \ref{comparison:a} --- \ref{comparison:e} depicts the impact of finding the most accurate initial states in the time execution. The sequential version (running in one thread) is the same for both algorithms, because they start the calculations from state $q_{0}$ on processing unit $P_{0}$. 
The Enumeration Approach requires more calculations for finite automata with larger number of states, larger input size and for higher number of processing units. 

The execution time of the Enumeration Approach compared to the PaREM algorithm increases as we increase either the input size or the number of threads. The execution time of PaREM is $2.3\times$ better than Enum, which is achieved in the largest number of threads (48) and the biggest problem size(1.07e+09), and only $1.04\times$  better than Enum for the smallest number of threads (6) and the smallest input size (6.69+e07).

\section{Related Work} \label{related_work}

Holub and Stekr \cite{HolubS09} propose an approach for parallel REM via DFA by splitting the input string in small chunks and running these chunks on each core, but due to pre-calculation of initial states for each sub input, this was not efficient for general DFA. Their algorithm runs efficiently for a specific type of DFA, so called synchronizing automata, that relies on the input automaton being k-local. 

Yang and Prassana \cite{yang2012} propose the segmentation of regular expressions and perform the REM evaluation via nondeterministic finite automata. The major aim is to optimize the use of memory hierarchy in case of automata with many states and large transition table. In contrast to our approach, the authors of \cite{yang2012} focus on large automata but do not address specifically algorithmic optimizations with respect to large input strings. 

Mytkowicz and Schulte \cite{mytkowicz2014} propose an approach that exploits SIMD, instruction and thread level parallelism in the context of finite state machines computations. To increase the opportunities for data-parallelism authors of [11] have devised a method for breaking data-dependencies with enumeration. This approach is not based on speculation with respect to initial state determination. 

Kumar et al. \cite{KumarCTV07} address the issue of large-scale finite automata (also known as the state explosion problem) by splitting regular expressions into two parts: (1) a prefix that contains frequently visited parts of the automata, and (2) a suffix that is the rest of the automaton. The aim is to have a small DFA for frequently accessed parts of automata that fits in cache memory. 

Luchaup et al. \cite{LuchaupSEJ11} propose an approach of finding the correct initial state by speculation. They believe that guessing the state of the DFA at certain position (network intrusion detection DFA based scanning spends most of the time in a few hot states) has a very good chance that after a few steps will reach the correct state. They validate these guesses using a history of speculated states. In comparison to our algorithm, the convergence of the guessed state and the correct state is not guaranteed. 
Furthermore, if a thread does not converge on its sub input, then the next thread is forced to start from a new state, which limits the scalability \cite{mytkowicz2014}.

Our algorithm is based on splitting the input into smaller sub-inputs (domain decomposition); however, we have devised a method to bypass the need of pre-calculation of all initial states by finding the most accurate possible initial states. Our approach is not limited to a particular type of DFA, and is efficient for a large spectrum of input sizes. 

In contrast to the related work, our tool is capable of automatically generating a ready to compile and execute code for shared-memory systems, by taking as input a RE or FA.

\section{Summary and Future Work} \label{summary_future_work}

Regular expression matching is essential for many applications such as lexical analysis, data mining \cite{Trasarti-data-mining}, or network security. We have presented a parallel algorithm for regular expression matching that is based on our improved speculative determination of initial states. 

Our tool PaREM transforms automatically any regular expression or finite automata into the corresponding parallel code (C++ and OpenMP), and consequently eases the access to the proposed parallel algorithm for the users without background in parallel programming. Preliminary experimental results show that the performance of our algorithm gracefully scales for various string lengths and numbers of threads. For an input string of 6.69e+07 characters, we obtained a speedup of $21\times$ with 48 threads.

In future, we plan to evaluate our approach for other types of problems, such as DNA sequencing or Network Intrusion Detection Systems. We also plan to extend our implementation for heterogeneous systems.

\bibliographystyle{IEEEtran}

\end{document}